\documentclass[12pt]{article}
\usepackage{amsmath}
\usepackage{amssymb}
\usepackage{cite}
\def\be{\begin{equation}}
\def\ee{\end{equation}}
\def\bea{\begin{eqnarray}}
\def\eea{\end{eqnarray}}

\def\bse{\begin{subequations}}
\def\ese{\end{subequations}}

\renewcommand\hat{\widehat}
\renewcommand\tilde{\widetilde}
\title{\bf Superstring partition functions in the doubled formalism}
\author{Sudipto Paul Chowdhury
\thanks{tpspc@mahendra.iacs.res.in}\\
\small
Department of Theoretical Physics\\
\small
Indian Association for the Cultivation of Science\\
\small
Calcutta - 700 032, India}
\date{}
\begin{document}
\maketitle
\begin{abstract}
\noindent Computation of superstring partition functions for the non-linear sigma
model on the product of a two-torus and its dual 
within the scope of the doubled formalism is presented.  We verify
that it reproduces the partition functions of the toroidally compactified
type--IIA and
type--IIB theories for appropriate choices of the GSO projection.
\end{abstract}
\thispagestyle{empty}
%%%
\section{Introduction} \label{intro}\label{sec0}
Duality symmetries of various kinds have proven to be extremely important in 
string theory. Generically, these relate different string theories.
In view of their usefulness, attempts have been made to 
formulate string theory in a duality-invariant manner resulting into M-theory 
\cite{ref1,ref2}, F-theory \cite{ref3,ref4,ref5}, 
S-theory \cite{ref6,ref7}. These formulations 
incorporated the geometric realization of dualities, based on geometric 
data. More recently, with the improved understanding of nongeometric 
backgrounds, attempts have been made to formulate a duality-invariant 
string theory incorporating non-geometric data in the scheme.
A formalism has been proposed, in which T-duality is made manifest 
by doubling the compact part of the target space \cite{ref8}. 
This formalism has been found to be consistent with 
many of the known results \cite{ref1,ref2,ref3,ref4,ref5,ref6,ref7,ref8,ref9,ref10,ref11,ref12,ref13,ref13a,ref13b,ref13c,ref13d,ref13e,ref13f,ref13g,ref13h,ref13i,ref13j}. Our calculations in the present
article provide further non-trivial support for the formalism,
extending to supersymmetric cases. This formulation,
extending the non-linear sigma model (NLSM), is referred to as the 
``doubled formalism". In this formulation  string theory is 
T-duality invariant and enhances spacetime dimensions by adding 
extra coordinates conjugate to  winding, called the dual coordinates. The 
bosonic string theory has been formulated in the so-called 
T-fold backgrounds with 
geometric constraints and it has been shown that upon adding a certain
topological term to the action, the corresponding quantum theory 
is equivalent to the quantum version of the non-linear sigma model defined on 
a worldsheet of arbitrary genus. A generalization of the formalism 
to superstring theories has also been worked out \cite{ref9} .
Constraint-quantization of the
doubled formalism has been studied too \cite{ref15,ref14}. 
In an attempt to relate results from the new theory with the usual results in
string theory, 
the partition function for the bosonic string on a circle 
has been calculated in the doubled formalism \cite{ref16}.
In the same vein it is important to compare the results for
superstrings and with targets with more than one compact dimensions.

Here we consider an $\mathcal{N} = (1,1)$ non-linear sigma model on a 
doubled torus $T^2 \times\hat{T}^2$, where the two-tori 
$T^2$ and $\hat{T}^2$ are 
dual to each other in the sense mentioned before. We compute the one-loop partition function of the two-torus.
%
%We find 
%that the partition function matches with the traditional type-II 
%results with a supersymmetry(susy)-extended topological term 
%added to the action. 
%
A two-torus is thought of as a direct product 
of circles, $T^2\simeq S^1\times S^1$. On each of the circles the superfields 
are split into ones with left and right chiralities. We  write down
the constraint equations for superfields on the doubled torus and find that 
they satisfy the appropriate chirality conditions. These constraints are
interpreted as the chiral superfields which is crucial
for establishing the quantum consistency of the doubled theory. 
A supersymmetrically extended topological term is needed for the superconformal
invariance of the theory. The bosonic part of the topological term
contributes with an overall sign factor to the partition function. The
fermionic part of the same, on the other hand, does not contribute.
This is of utmost importance for the matching of the partition
function with the type-II results.

In section~\ref{sec1} we write down the action of
the doubled $\mathcal{N}=(1,1)$ NLSM as well as the superfields along
with the constraints.
The equation of motion for bosons on a torus have instanton solutions.
In section~\ref{sec2} we present the computation of the one-loop partition
function for the instanton 
sector for bosons \cite{ref16} on a two-torus.
A Poisson re-summation is required for the holomorphic
factorization of the partition function. These computations for the bosons
yield the sum over the internal momenta. We discuss the contributions from 
the bosonic and fermionic oscillators to the partition function,
in section~\ref{sec3}, in terms of the well-known modular functions.
The fermionic contributions to the partition function after suitable
GSO projections are found to match with the type-IIA and type-IIB results. Finally in section \ref{sec4} we draw conclusions from our work.
%%%
\section{The $\mathcal{N} =(1,1)$ NLSM on a doubled torus}
\label{sec1}
Let us  start with the non-linear sigma model action in $N=(1,1)$ superspace 
on a doubled torus, $T^2 \times \hat T^2$, where $T^2$ and $\hat T^2$ are 
dual to each other in the sense that the torus $T^2$ parametrizes the
compact part of the 10 dimensional target space, while the torus $\hat{T}^2$
parametrizes the directions conjugate to the windings. In this article we 
calculate the partition function solely on the compact part of the target 
space, that is the torus $T^2$ and its dual.
The torus $T^2$ has radii $R_1$and $R_2$ for its two circles, while the 
dual torus $\hat{T}^2$ has radii $1/R_1$ and $1/R_2$, respectively.
Unhatted and hatted expressions are used to define quantities on 
$T^2$ and $\hat{T}^2$, respectively. For example, we denote the superfields on $T^2$
and $\hat{T}^2$ by $\Phi$ and $\hat{\Phi}$, respectively.
The $\mathcal{N} = (1,1)$ non-linear sigma model action generalizing the corresponding bosonic action is
\begin{equation} \label{phact}
S = \frac{\pi}{4}\int{d^2}zd^2\theta[g_{\mu \nu}\epsilon^{ab} D_a\Phi^\mu
D_b\Phi^\nu
+ \hat{g}_{\mu \nu}\epsilon^{ab} D_a\hat{\Phi} ^\mu
D_b\hat{\Phi}^\nu]. \end{equation}
The superfields are
functions of $(x,\theta,\overline{\theta})$, where $x$ represents the 
spacetime coordinates and $\theta$ and $\overline{\theta}$ are the 
mutually conjugate Grassmannian supercoordinates. The superfields are 
expanded in terms of scalars, Majorana-Weyl spinors 
and the auxiliary fields as \cite{ref17}
\begin{gather}\label{phexp}
\Phi^\mu(x,\theta, \overline{\theta}) = X^\mu(x) +  i\theta\psi^\mu(x)+
i\overline{\theta}\tilde\psi^\mu(x) +
\theta\overline{\theta}F^\mu(x),\\
\label{phexp2}
 \hat{\Phi}^\mu(x,\theta, \overline{\theta}) = \hat{X}^\mu(x) +
i\theta\hat{\psi}^\mu(x)+
 i\overline{\theta}\tilde{\hat{\psi}}^{\mu}(x) +
\theta\overline{\theta}\hat{F}^\mu(x).
\end{gather}
In (\ref{phact}), the measure over the Grassmann coordinates is defined as
\begin{equation}\label{deriv}
d^2 \theta = d\theta d\overline{\theta},
\end{equation}
where the targetspace indices are denoted by 
$\mu,\nu = 1, 2$ and the worldsheet indices by $ a,b = 1, 2$.
The supercovariant derivatives are defined as
\begin{gather}\label{supder}
 D_1 = D_\theta = \frac {\partial} {\partial \theta} + \theta \frac{\partial}
{\partial z},
\\\label{supder2}
D_2 = D_ {\overline{\theta}} = \frac{\partial} {\partial \overline{\theta}} +
\overline{\theta} \frac{\partial} {\partial \overline{z}}.
\end{gather}
where $z$ denote the complex coordinate of the Euclidean worldsheet,
while  $\overline{z}$ denotes its complex conjugate. The metric tensor for
the tori $T^2$ and $\hat{T}^2$ are
\begin{gather}\label{metr}
g_{\mu \nu} = \left(\begin{array}{cc}{R_1}^2 & 0\\0&{R_2}^2\end{array}\right),
\\
\label{metr2}
 \hat{g}_{\mu \nu} = \left(\begin{array}{cc}{R_1}^{-2} &
0\\0&{R_2}^{-2}\end{array}\right),
\end{gather}
respectively.

Using the expressions (\ref{phact}) through (\ref{metr2}) the action $S$
assumes the form
\begin{equation} \label{xact}
\begin{split}
S = \frac{\pi}{2}\int d^2z &\left[{R_1}^2(\partial_{\overline{z}}X^1\partial_z X_1
+ \psi^1\partial_{\overline{z}}\psi_1 + \tilde{\psi}^1\partial_z\tilde{\psi}_1
+ F^1
F_1) \right.\\
&+ {R_2}^2(\partial_{\overline{z}}X^2\partial_z X_2 +
\psi^2\partial_{\overline{z}}\psi_2 +
\tilde{\psi}^2\partial_z{\tilde{\psi}}_2 + F^2F_2) \\
&+ {R_1}^{-2}(\partial_{\overline{z}}\hat{X}^1\partial_z \hat{X}_1 +
\hat{\psi}^1\partial_{\overline{z}}\hat{\psi}_1 +
\tilde{\hat{\psi}}^1\partial_z\tilde{\hat{\psi}}_1 + \hat{F}^1 \hat{F}_1) \\
&+ \left.{R_2}^{-2}(\partial_{\overline{z}}\hat{X}^2\partial_z \hat{X}_2 +
\hat{\psi}^2\partial_{\overline{z}}\hat{\psi}_2 +
\tilde{\hat{\psi}}^2\partial_z\tilde{\hat{\psi}}_2 + \hat{F}^2
\hat{F}_2)\right].
\end{split}
\end{equation}
This action is not invariant under the T-duality transformations :
\begin{equation}\label{tdual}
R_i \rightarrow \frac{1}{R_i}, \quad i=1,2.
\end{equation}
To make it T-duality invariant we introduce the 
geometric constraint equations as in \cite{ref16}. 
Towards this we introduce the superfields 
\begin{gather}\label{constr1}
\textbf{P}^1(x, \theta, \overline{\theta}) = R_1\Phi^1 + {R_1}^{-1}
\hat{\Phi}^1,
\\
\label{constr2}
 \textbf{P}^2(x, \theta, \overline{\theta}) = R_2\Phi^2 + {R_2}^{-1}
\hat{\Phi}^2,
\\
\label{constr3}
\textbf{Q}^1(x, \theta, \overline{\theta}) = R_1\Phi^1 - {R_1}^{-1}
\hat{\Phi}^1,
\\
\label{constr4}
\textbf{Q}^2(x, \theta, \overline{\theta}) = R_2\Phi^2 - {R_2}^{-1}
\hat{\Phi}^2.
\end{gather}
\textbf{P} and \textbf{Q} can be expanded in terms of components, such
as scalars $P^{\mu}$, $Q^{\mu}$ and Majorana-Weyl spinors $\psi_p$'s and 
$\psi_q$'s. 
\begin{gather}\label{consex}
\textbf{P}^\mu = P^\mu + i \theta {\psi_p}^\mu +
i\overline{\theta}\tilde{\psi_p}^\mu
+ \theta \overline{\theta}{F_p}^\mu,
\\
\label{consex2}
\textbf{Q}^\mu = Q^\mu + i \theta {\psi_q}^\mu +
i\overline{\theta}\tilde{\psi_q}^\mu
+ \theta \overline{\theta}{F_q}^\mu,
\end{gather}
where $\mu,\nu=1,2$, and 
\begin{alignat}{3}
\label{pscal}
 P^1 = R_1 X^1 + R_1^{-1} \hat{X}^1,
&\quad
%\label{psip}
 {\psi_p}^1 = R_1 \psi^1 + R_1^{-1}\hat{\psi}^1,
&\quad
%\label{psipt}
\tilde{\psi_p}^1 = R_1 \psi^1 + R_1^{-1}\hat{\psi}^1,
\\
\label{pscal2}
%\label{qscal}
 Q^1 = R_1 X^1 - R_1^{-1}\hat{X}^1,
&\quad
%\label{psiq}
 {\psi_q}^1 = R_1 \psi^1 - R_1^{-1}\hat{\psi}^1,
&\quad
%\label{psiqt}
\tilde{\psi}_q^1 = R_1\tilde{\psi}^1 - R_1^{-1}\tilde{\hat{\psi}^1},
\end{alignat}
where $i=1,2$.
A topological term containing the superfields is also added to the 
action (\ref{phact}) to
ensure invariance under large gauge transformations corresponding to
the holomorphic and anti-holomorphic $U(1)$ currents
on $T^2\simeq S^1\times S^1$, generalizing the topological term for
the bosonic case \cite{ref16}:
\begin{equation}\label{topact}
S_{top} = \pi \int d^2z d^2\theta [D_{\overline{\theta}}\Phi^1 D_\theta
\hat{\Phi}^1 +
D_\theta\Phi^1 D_{\overline{\theta}} \hat\Phi^1 + D_{\overline{\theta}}\Phi^2
D_\theta \hat{\Phi}^2 + D_\theta\Phi^2 D_{\overline{\theta}}\hat{\Phi}^2].
\end{equation}
The superfields \textbf{P} and \textbf{Q} are subject to the chirality
constraints
\begin{gather}\label{chiral}
 \int d^2\theta D_{\overline{\theta}}\textbf{P}^\mu = 0,
\\
\int d^2\theta D_{\theta}
\textbf{Q}^\mu = 0, 
\end{gather}
thus making \textbf{P} and \textbf{Q} holomorphic and anti-holomorphic, 
respectively.
Using the expressions (\ref{pscal}) and (\ref{pscal2}), 
the action (\ref{xact}) becomes  
\begin{equation}\label{dbactp}
\begin{split}
S = \frac{\pi}{4} \int d^2z &[(\partial_{\overline{z}}P^1\partial_z P^1
+ {\psi_p}^1\partial_{\overline{z}}{\psi_p}^1 +
\tilde{\psi_p}^1\partial_z\tilde{\psi_p}^1) \\ 
&+ (\partial_{\overline{z}}P^2\partial_z P^2 +
{\psi_p}^2\partial_{\overline{z}}{\psi_p}^2 +
\tilde{\psi_p}^2\partial_z{\tilde{\psi_p}}^2)  \\ 
&+ (\partial_{\overline{z}}Q^1\partial_z {Q}^1 +
{\psi_q}^1\partial_{\overline{z}}{\psi_q}^1 +
\tilde{\psi_q}^1\partial_z\tilde{\psi_q}^1) \\ 
&+ (\partial_{\overline{z}}{Q}^2 \partial_z {Q}^2 +
{\psi_q}^2 \partial_{\overline{z}}{\psi_q}^2 +
{\tilde{\psi_q}}^2\partial_z\tilde{\psi_q}^2)].
\end{split}
\end{equation}
This action has no explicit dependence on the radii of the two
two-tori and is thus manifestly T-duality invariant. 
The action (\ref{dbactp}) is quantum equivalent to that 
of superstring theory only if the one-loop partition functions of both
theories match.
This necessitated the addition of a topological term in the bosonic
case. We extend the topological term by incorporating the
corresponding fermionic contributions so as to preserve $\mathcal{N} = (1,1)$ superconformal symmetry. The extended topological
term takes the form 
\begin{equation}\label{topact2}
\begin{split}
S_{top} = &\pi \int d^2 z[-\frac{1}{4}(\partial_{\overline{z}}P^1 \partial_z Q^1 - 
\partial_z P^1\partial_{\overline{z}} Q^1 + \partial_{\overline{z}}P^2 \partial_z Q^2 - 
\partial_z P^2 \partial_{\overline{z}} Q^2) \\
&+ \frac{1}{2}(\psi^1_p \partial_{\overline{z}} \psi^1_q - \psi^1_q \partial_{\overline{z}} \psi^1_p + \psi^2_p \partial_{\overline{z}} \psi^2_q - \psi^2_q \partial_{\overline{z}} \psi^2_p) \\
&+ \frac{1}{2}(\tilde{\psi}^1_p \partial_z \tilde{\psi}^1_q - \tilde{\psi}^1_q \partial_z \tilde{\psi}^1_p + \tilde{\psi}^2_p \partial_z \tilde{\psi}^2_q - \tilde{\psi}^2_q \partial_z \tilde{\psi}^2_p)].
\end{split}
\end{equation}
The equations of motion for the scalars 
obtained from the action (\ref{xact}) have instanton solutions. These
solutions are classical and come from periodicity conditions on the 
compact bosons. They contribute to the partition function with the
sums over the bosonic momenta, to which we now turn.
\section{The Instanton Contributions}
\label{sec2}
In this section we consider the contribution of the bosonic instanton sector 
to the partition function. To calculate the partition function for
chiral bosons one needs to employ the holomorphic factorization technique 
so as to retain the contribution with the right holomorphic dependence. 
We apply here the same technique for $P$ and $Q$ as in \cite{ref16}. 

In calculating the partition function, the superfields $\Phi$ 
in the action (\ref{phact}) are replaced by the combinations
$\textbf{L}$ and $\hat{\textbf{L}}$, periodic under
the shifts in the momenta of the scalars along the circles of $T^2$.
These are given by 
\begin{gather}\label{phiel1}
\textbf{L}^\mu = \int D_\theta {\Phi}^\mu dz d \theta + \int
D_{\overline{\theta}}{\Phi}^\mu d{\overline{z}} d{\overline{\theta}} +
N\alpha^\mu + M\beta^\mu + N^\prime\alpha^\mu + M^\prime\beta^\mu,
\\
\label{phiel2}
\hat{\textbf{L}}^\mu = \int D_\theta \hat{\Phi}^\mu dz d\theta + \int
D_{\overline{\theta}} \hat{\Phi}^\mu d{\overline{z}} d{\overline{\theta}} +
\hat{N}\alpha^\mu + \hat{M}\beta^\mu + \hat{N}^\prime\alpha^\mu +
\hat{M}^\prime\beta^\mu,
\end{gather}
where $\alpha$ and $\beta$ designates the 1-cycles of the tori,
corresponding to $T^2\simeq S^1\times S^1$.
The bosonic parts of these combinations are
\begin{gather}\label{elb}
 {L^\mu}_b = \int d{X}^\mu + N\alpha^\mu + M\beta^\mu,
\\
\label{elb2}
{\hat{L}^\mu}_b = \int d\hat{X}^\mu + \hat{N} \alpha^\mu + \hat{M}\beta^\mu,
\end{gather}
where
\begin{equation}\label{matxn}
N = \left(\begin{array}{cc}{n}_1 & 0\\0&{n}_2\end{array}\right),\quad \hat{N}
=\left(\begin{array}{cc}{\hat{n}}_1 & 0\\0&{\hat{n}}_2\end{array}\right),
\end{equation}
and
\begin{equation}\label{matxm}
M = \left(\begin{array}{cc}{m}_1 & 0\\0&{m}_2\end{array}\right),\quad\hat{M} =
\left(\begin{array}{cc}{\hat{m}}_1 & 0\\0&{\hat{m}}_2\end{array}\right).
\end{equation}
In equations (\ref{elb}) and (\ref{elb2}) 
\begin{equation}\label{dxex}
dX = \partial_z X dz + \partial_{\overline{z}}X d{\overline{z}},\quad d{\hat X} = \partial_z \hat X dz + \partial_{\overline{z}}\hat X d{\overline{z}}
 \end{equation}
The fermionic parts of $\mathbf{L}$ and $\hat{\mathbf{L}}$ are given by 
\begin{gather}\label{elf1}
 L^\mu_f = i \int dz d\theta [{\psi}^\mu_p +
\theta\overline{\theta}\partial_z{\tilde{\psi}^\mu _p}] + i \int
d{\overline{z}} d{\overline{\theta}}[\tilde{\psi}^\mu_p +
\overline{\theta}\theta \partial_{\overline{z}}{\psi}^\mu _p],
\\
\label{elf2}
\hat{L}^\mu_f = i \int dz d\theta [\hat{\psi}^\mu_p +
\theta \overline{\theta} \partial_z {\tilde{\hat{\psi}}^\mu _p}] + i \int
d{\overline{z}} d{\overline{\theta}} [\tilde{\hat{\psi}}^\mu_p +
\overline{\theta}\theta \partial_{\overline{z}}{\hat{\psi}}^\mu _p].
\end{gather}
Similarly, the superfields $\mathbf{P}$ and $\mathbf{Q}$ are also to 
be combined into
periodic combinations with bosonic and fermionic parts as
\begin{gather}\label{instm}
{\Psi}^\mu_b = \int d{P}^\mu + (RN + R^{-1}\hat{N})\alpha^\mu + (RM +
R^{-1}\hat{M})\beta^\mu,
\\
\label{instn}
{\Upsilon}^\mu_b = \int d{Q}^\mu + (RN - R^{-1}\hat{N})\alpha^\mu + (RM -
R^{-1}\hat{M})\beta^\mu.
\end{gather}
and the fermions $\psi_p$ and $\psi_q$ in the same equations are replaced by
\begin{gather}\label{instfm}
{\Psi}^\mu_f = i\int dz d\theta [{{\psi}^\mu_p} + \theta
{\overline{\theta}}\partial{z}\tilde{\psi}^\mu_p] + i\int d{\overline{z}}
d{\overline{\theta}}[{{\tilde{\psi}^\mu}_p}
+{\overline{\theta}}\theta\partial_{\overline{z}}\psi^\mu_p],
\\
\label{instfn}
{\Upsilon}^\mu_f = i\int dz d\theta [{{\psi}^\mu_q} + \theta
{\overline{\theta}}\partial{z}\tilde{\psi}^\mu_q] + i\int d{\overline{z}}
d{\overline{\theta}}[{{\tilde{\psi}^\mu}_q}
+{\overline{\theta}}\theta\partial_{\overline{z}}\psi^\mu_q].
\end{gather}
Here
\begin{gather}\label{matrixr1}
 R  = \left(\begin{array}{cc}{R}_1 & 0\\0&{R}_2\end{array}\right),
\\
\label{matrixr2}
 R^{-1}  = \left(\begin{array}{cc}\frac{1}{R_1} & 0\\0&\frac{1}{R_2}\end{array}\right).
\end{gather}
Rewriting the action (\ref{dbactp}) in terms of the periodic
combinations, ${\Psi}^\mu_b$'s and ${\Upsilon}^\mu_b$'s, and using equations 
(\ref{elf1})---(\ref{instn}), one can extract the terms independent of ${\Psi}^\mu_b$'s and ${\Upsilon}^\mu_b$'s. These terms contribute to the ``instanton'' sum. The instanton sector of the partition function contains sum over
all field configurations,
\begin{equation}\label{partb}
\begin{split}
Z^{inst}_b = &\sum_{{n_1,m_1},\atop{\hat{n}_1,\hat{m}_1}}
\sum_{n_2,m_2,\atop{\hat{n}_2,\hat{m}_2}}
\exp(
{-(R_1n_1 + R^{-1}_1\hat{n}_1)^2\frac{\pi{|\tau^1|}^2} {4\tau^1_2}}\\
&+ {(R_1n_1 + R^{-1}_1\hat{n}_1)(R_1m_1 + R^{-1}_1\hat{m}_1)
\frac{\pi \tau^1_1}{2\tau^1_2}}\\
&-{(R_1m_1 + R^{-1}_1\hat{m}_1)^2\frac{\pi}{4\tau^1_2}}
-{(R_2n_2 + R^{-1}_2\hat{n}_2)^2\frac{\pi{|\tau^2|}^2} {4\tau^2_2}} \\
& + (R_2n_2 + R^{-1}_2\hat{n}_2)(R_2m_2 + R^{-1}_2\hat{m}_2)\frac{\pi
\tau^2_1}{2\tau^2_2} - (R_2m_2 + R^{-1}_2\hat{m}_2)^2\frac{\pi}{4\tau^2_2}\\
&-(R_1n_1 - R^{-1}_1\hat{n}_1)^2\frac{\pi{|\tau^1|}^2} {4\tau^1_2}
 +(R_1n_1 - R^{-1}_1\hat{n}_1)(R_1m_1 - R^{-1}_1\hat{m}_1)\frac{\pi
\tau^1_1}{2\tau^1_2}\\
&-(R_1m_1 - R^{-1}_1\hat{m}_1)^2\frac{\pi}{4\tau^1_2}
-(R_2n_2 - R^{-1}_2\hat{n}_2)^2\frac{\pi{|\tau^2|}^2} {4\tau^2_2}\\
& +(R_2n_2 - R^{-1}_2\hat{n}_2)(R_2m_2 - R^{-1}_2\hat{m}_2)\frac{\pi
\tau^2_1}{2\tau^2_2} 
-(R_2m_2 - R^{-1}_2\hat{m}_2)^2\frac{\pi}{4\tau^2_2}).
\end{split}
\end{equation}
Here $\tau^1_1$ and $\tau^1_2$ are respectively the real and complex
parts of the modular parameter of the torus $T^2$ 
while $\tau^2_1$ and $\tau^2_2$ are their counterparts for the dual torus $\hat T^2$.
The fermionic parts of the periodic combinations, ${\Psi}^\mu_f$ and ${\Upsilon}^\mu_f$ do not couple with the momenta and windings and hence do not contribute to the partition function. They simply reproduce the classical action (\ref{xact}) after being squared. The bosons in the topological term contribute only a sign 
\begin{equation}\label{toppartb}
 Z^{top}_b = \prod_{i} \exp[i\pi (n_i \hat m_i - m_i \hat n_i)]
\end{equation}
to the partition function, as mentioned above.

Holomorphic factorization of the partition function calls for Poisson
re-summation. For that one first has to separate the
contributions from the scalars $P$'s and $Q$'s in terms of independent variables.
Let us consider the case where the radii of the tori are
$R^2_i = \frac{\rho_i}{\lambda_i}$ with coprime integers
$\rho_i$ and $\lambda_i$ and let us
define $\xi_i = {\rho_i}{\lambda_i}$, $i=1,2$.
Then \cite{ref16}
\begin{equation}\label{posrs1}
R_i n_i \pm R^{-1}_i\hat{n}_i = \sqrt{\xi_i}(\frac{n_i}{\lambda_i} \pm
\frac{\hat{n}_i}{\rho_i}).
\end{equation}
Now we make a substitution
\begin{equation}\label{posrs2}
\begin{split}
n_i &= c_i\lambda_i + \lambda_i \sigma_{\lambda_i} \\
\hat{n}_i &= \hat{c}_i\rho_i + \rho_i \sigma_{\rho_i},
\end{split}
\end{equation}
where
\begin{equation}\label{posrs3}
c_i,\hat{c_i}\in 
\mathbb{Z}\text{~and~}\sigma_{\lambda_i}\in \{0,1/{\lambda_i},\cdots,{\lambda_i-1}/\lambda_i\}.
\end{equation}
We can write
\begin{equation}\label{posrs4}
 \sqrt{\xi_i}(\frac{n_i}{\lambda_i} \pm \frac{\hat{n}_i}{\rho_i}) = \sqrt{\xi_i}(c_i \pm
\hat {c_i}
+ {\sigma}_{\lambda_i} \pm {\sigma}_{\rho_i}).
\end{equation}
A further substitution with $h_i = c_i + \hat{c}_i$ and 
$l_i = c_i - \hat{c}_i$ allows us to rewrite the sum over 
$n_i$ and $\hat{n}_i$ as sum over $h_i$ and $l_i$ $\in \mathbb{Z}$. 
Since $h_i - l_i = 2\hat{c}_i$, we restrict to even values of $h_i - l_i$.
This is done by inserting a factor of
\begin{equation*} 
\frac{1}{2} \sum_{\phi \in \{0,1/2\}} \exp{\left[2\pi i \phi (h_i - l_i)\right]}
\end{equation*} 
in (\ref{partb}). One can repeat the process for the $m_i$ and $\hat{m}_i$ sums and including the contribution from the bosonic parts of the topological terms, the instanton piece of the partition function becomes
\begin{equation}\label{partb1}
\begin{split}
Z^{inst}_b =& \prod_{i}\sum_{\phi,\chi,\sigma_{\lambda_i},\sigma_{\rho_i},\sigma^{\prime}_
{\lambda_i},\sigma^{\prime}_{\rho_i}\atop{h_i,l_i,s_i,t_i}} 
\frac{1}{16}
\exp[
-\frac{\pi \xi_i}{4}((h_i + \sigma_{\lambda_i}+\sigma_{\rho_i})^2 \frac{|\tau^i|^2}{\tau^i_2}\\
- &2(h_i + \sigma_{\lambda_i} + \sigma_{\rho_i})(s_i + \sigma^{\prime}_{\lambda_i} + \sigma^{\prime}_{\rho_i})\frac{\tau^i_1}{\tau^i_2}
+ (s_i + \sigma^{\prime}_{\lambda_i} + \sigma^{\prime}_{\rho_i})^2 \frac{1}{\tau^i_2}\\ &+ (l_i + \sigma_{\lambda_i} - \sigma_{\rho_i})^2 \frac{|\tau^i|^2}{\tau^i_2} - 2(l_i + \sigma_{\lambda_i} - \sigma_{\rho_i} )(t_i + \sigma^{\prime}_{\lambda_i} - \sigma^{\prime}_{\rho_i})\frac{\tau^i_1}{\tau^i_2}\\
&+ (t_i + \sigma^{\prime}_{\lambda_i} - \sigma^{\prime}_{\rho_i})^2\frac{1}{\tau^i_2}) + 2\pi i (\phi(h_i - l_i) + \chi(s_i - t_i))\\
&+ \frac{i\pi\xi}{2}((l_i + \sigma_{\lambda_i} - \sigma_{\rho_i})(s_i + \sigma'_{\lambda_i} + \sigma'_{\rho_i}) - (h_i + \sigma_{\lambda_i} + \sigma_{\rho_i})(t_i + \sigma'_{\lambda_i} - \sigma'_{\rho_i})].
\end{split}
\end{equation}

Now we define $\sigma^{\pm}_i = \sigma _{\lambda_i} \pm \sigma _{\rho_i}$. After summing over $s_i$ and $t_i$, the contribution from the scalar fields $P$ to the partition function (\ref{partb1}) is given by
\begin{equation}\label{partb2}
\begin{split}
 Z^{instP}_b = & \prod_i
\sum_{\phi,\chi,\sigma_{\lambda_i},\sigma_{\rho_i},\sigma^{\prime}_
{\lambda_i},\sigma^{\prime}_{\rho_i}\atop{h_i,l_i,u_i}}\frac{1}{8}\sqrt{\frac{4\tau^i_2}{\xi_i}}
\exp[-\frac{\pi\xi_i}{4}((h_i + \sigma^{+}_i)^2\frac{|\tau^i|^2}{\tau^i_2}\\
& - 2\sigma'^{+}_i(h_i + \sigma^{+}_i)\frac{\tau^i_1}{\tau^i_2} + (\sigma'^{+}_i)^2 \frac{1}{\tau^i_2})
+ 2\pi i \phi h_i + \frac{i\pi \xi_i}{2}(l_i + \sigma^{-}_i)\sigma'^{+}_i \\
&- \frac{4\pi\tau^i_2}{\xi_i}(u_i - \chi + i\xi_i \frac{(h_i + \sigma^{+}_i)}{4}\frac{\tau^i_1}{\tau^i_2} - \frac{i\xi_i\sigma'^{+}_i}{4\tau^i_2} - \frac{\xi_i}{4}(l_i + \sigma^{-}_i)^2)].
\end{split}
 \end{equation}
On rearranging the sum  takes the form
\begin{equation}\label{partb3}
\begin{split}
Z^{instP}_b =
&\prod_i \sum_{\phi,\chi,\sigma_{\lambda_i},\sigma_{\rho_i},\sigma^{\prime}_
{\lambda_i},\sigma^{\prime}_{\rho_i}\atop{h_i,l_i,u_i}}\sqrt{\frac{\tau^i_2}{\xi_i}}
\exp[
-\frac{\pi\xi_i}{4}((h_i + \sigma^{+}_i)^2 \\
&- 4\xi_i(\frac{u_i - \chi}{\xi_i}
 - \frac{1}{4}(l_i + \sigma^{-}_i) )^2
+ \frac{i\pi \xi_i}{2}(l_i + \sigma^{-}_i)\sigma'^{+}_i \\
&- 2\pi\tau^i_1(h_i + \sigma^{+}_i)(u_i - \chi - \frac{\xi_i}{4}(l_i + \sigma^{-}_i)) + 2\pi i \phi h_i + 2\pi i (u_i - \chi)\sigma'^{+}_i].
\end{split}
\end{equation}
Similar expressions can be written for the fields $Q$ with $h_i$ replaced by $l_i$. Combining the contributions from the scalar fields $P$'s and $Q$'s, the instanton part of the partition function for the bosons is written as
\begin{equation}\label{partb4}
\begin{split}
 Z^{inst}_b = & \prod_i
\sum_{\phi,\chi,\sigma_{\lambda_i},\sigma_{\rho_i},\sigma^{\prime}_
{\lambda_i},\sigma'_{\rho_i}\atop{h_i,l_i,u_i,\hat{u}_i}}(\sqrt{\frac{\tau^i_2}{2\xi_i}} \exp[
i\pi\xi_i\tau^i\frac{{p^i_L}^2}{2} - i\pi\xi_i\overline{\tau}^i\frac{{{p^i_R}^2}}{2}\\
& + 2\pi i(\phi h_i + (u_i - \chi)\sigma'^{+}_i)])\\
&\times(\sqrt{\frac{\tau^i_2}{2\xi_i}}
\exp[i\pi\xi_i\tau^i\frac{{q^i_L}^2}{2} - i\pi\xi_i\overline{\tau}^i\frac{{{q^i_R}^2}}{2} + 2\pi i(\phi l_i + (\hat{u}_i + \chi)\sigma'^{+}_i)]),
\end{split}
\end{equation}
where
\begin{equation}\label{posrs6}
\begin{split}
p^i_L = \frac{1}{2}(h_i + \sigma^{+}_i) - 2(\frac{u_i - \chi}{\xi_i} - \frac{1}{4}(l_i
+\sigma^{-}_i)),
\\
p^i_R = \frac{1}{2}(h_i + \sigma^{+}_i) + 2(\frac{u_i - \chi}{\xi_i} - \frac{1}{4}(l_i +\sigma^{-}_i)),
\\
q^i_L = \frac{1}{2}(l_i + \sigma^{-}_i) - 2(\frac{\hat{u}_i + \chi}{\xi_i} - \frac{1}{4}(h_i
+\sigma^{+}_i)),
\\
q^i_R = \frac{1}{2}(l_i + \sigma^{-}_i) + 2(\frac{\hat{u}_i + \chi}{\xi_i} - \frac{1}{4}(h_i
+\sigma^{+}_i)).
\end{split}
\end{equation}
The sums over $h_i$, $l_i$, $\sigma^{+}_i$, $\sigma^{-}_i$, $\phi$ etc can be replaced by sums over $n_i$ and $\hat{n}_i$ using the expressions (\ref{posrs2}) and making use of the identity
\begin{equation}\label{posrs7}
 \sum_{\xi=0}\left(\exp{\left(\frac{2\pi i \xi}{n}\right)}\right)^j =
\sum_{\sigma_n}\exp{\left(2\pi i {\sigma_n}j\right)} = 
\begin{cases}
 n, \text{if~} j = 0 \mod n \\ 
 0, \text{~otherwise.}
\end{cases}
\end{equation}
Consequently, we have
\begin{equation}\label{posrs8}
\begin{split}
&u_i + \hat{u}_i = 0\mod \lambda_i, \\
&u_i - \hat{u}_i - 2\chi = 0\mod \rho_i,
\end{split}
\end{equation}
and these criteria are satisfied by the choice
\begin{equation}\label{posrs9}
 \frac{u_i - \chi}{\xi_i} = \frac{1}{2}\left(\frac{{\omega}_i}{{\rho}_i} + 
\frac{\hat{\omega}_i}{\lambda_i}\right),\quad \frac{\hat{u}_i + \chi}{\xi_i} = \frac{1}{2}\left(\frac{{\omega}_i}{{\rho}_i} - 
\frac{\hat{\omega}_i}{\lambda_i}\right).
\end{equation}
where we have replaced the sums over $u_i, \hat{u}_i, \chi, \sigma'_{\lambda_i}~ \mbox{and}~    \sigma'_{\rho_i}$ by sums over $w_i~ \mbox{and}~ \hat{w}_i \in \mathbb{Z}$. We can now identify the left-moving and right-moving momenta as
\begin{equation}\label{posrs10}
\begin{split}
p^i_L &= \frac{n_i}{\lambda_i} - \left(\frac{{\omega}_i}{{\rho}_i} +
\frac{\hat{\omega}_i}{\lambda_i}\right). \\
p^i_R &= \frac{\hat{n}_i}{\lambda_i} + \left(\frac{{\omega}_i}{{\rho}_i} +
\frac{\hat{\omega}_i}{\lambda_i}\right). \\
q^i_L &= - \frac{\hat{n}_i}{\rho_i} - \left(\frac{{\omega}_i}{{\rho}_i} -
\frac{\hat{\omega}_i}{\lambda_i}\right) .\\
q^i_R &=  \frac{n_i}{\rho_i} + \left(\frac{{\omega}_i}{{\rho}_i} -
\frac{\hat{\omega}_i}{\lambda_i}\right).
\end{split}
\end{equation}
The doubled partition function for the bosons can now be written as
\begin{equation}\label{partd}
\begin{split}
Z^{doubled}_b =& \prod_i \sum_{p^i_L,p^i_R}\sqrt{2\tau^i_2}\exp{\left[i\pi \xi_i
\tau^i\frac{{p^i_L}^2}{4} - i\pi \xi_i\overline{\tau}^i\frac{{p^i_R}^2}{4}\right]}\\
&\times~ \sum_{q^i_L,q^i_R}\sqrt{2\tau^i_2}\exp{\left[i\pi \xi_i
\tau^i\frac{{q^i_L}^2}{4} - i\pi \xi_i\overline{\tau}^i\frac{{q^i_R}^2}{4}\right]}.
\end{split}
\end{equation}
Since the momentum sums are decoupled we can keep only the holomorphic part of the doubled partition function from both $P$s and $Q$s. This gives us the contribution of the bosonic fields to the partition functions on a two-torus.
\begin{equation}\label{parth}
Z^{holo}_b = \prod_i \sum_{p^i_L,q^i_R}\sqrt{2\tau^i_2}\exp{\left[i\pi \xi_i 
\tau^i\frac{{p^i_L}^2}{4} - i\pi \xi_i\overline{\tau}^i\frac{{q^i_R}^2}{4}\right]}.
\end{equation}

 The holomorphic factorization of the ``instanton'' sum guarantees the inclusion of all 
spin structures necessary for the chiral bosons.
Thus the contributions from the bosonic instanton sector to the partition function give us the sum over the entire momentum lattice.
%%%
\section{The Oscillator Contributions}
\label{sec3}
The contribution to the partition function
from the oscillator sectors of bosons and fermions can
be obtained by evaluating the 
path integral with the action (\ref{xact}). For the bosonic case this yields the (squared)
partition function of the bosonic string theory \cite{ref16} on a two-torus.
Let us discuss this in brief. The path integral for bosons  is
\begin{equation}\label{partos}
Z^{osc}_b = \int DX^1 DX^2 \exp{\left[-\frac{\pi}{2} \int d^2
z[R^2_1\partial_{\overline{z}}X^1\partial_{z}X^1
+ R^2_2\partial_{\overline{z}}X^2\partial_{z}X^2]\right]},
\end{equation}
which evaluates to 
\begin{equation}\label{partos1}
Z^{osc}_b = \frac{R_1}{\sqrt{2{\det \Box}}} \frac{R_2}{
\sqrt{2{\det \Box}}},\end{equation}
with 
$\det \Box  =  \tau^1_2 \eta^2(\tau) \eta^2(\overline{\tau})$,
where $\eta(\tau)$ is the Dedekind $\eta$-function
\begin{equation}\label{eta}
\eta(\tau) = e^{i\pi/12}\prod_{n >1}(1-e^{2\pi in\tau}).
\end{equation}
where $\tau$ is the modular parameter on the worldsheet torus.

The path integral over the dual fields $\hat{\phi}$ gives 
a similar expression, the only difference being that the radii
$R_1$ and $R_2$ appear in the denominator. Taking all the bosonic
contributions into account, the square-root of the path integral
contributes the following piece to the partition function
\begin{equation}\label{partfb}
Z_{b} = \prod_i\sum_{p^i_L,q^i_R}\frac{1}{|\eta|^4} \exp{\left[i\pi\tau\frac{{(p^i_L)}^2}{4}-
i\pi\overline{\tau}\frac{{(q^i_R)}^2}{4}\right]}.
\end{equation}

Since the radii of the two-torus appearing in the expression (\ref{partos1}) cancel with those appearing in the expression for the dual bosons the final expression above for the bosons is T-duality invariant. Let us now discuss the fermionic contributions.
The path integral for the fermions is
\begin{equation}\label{partosf2}
\begin{split}
Z^{osc}_f = 
 \int D\psi^1 D\psi^2 D\tilde{\psi}^1 D\tilde{\psi}^2
&\exp{\Big(-\frac{\pi}{2} \int d^2z \big( 
R^2_1(\psi^1\partial_{\overline{z}}\psi_1 +
\tilde{\psi}^1\partial_z\tilde{\psi}_1)} \\
&+ {R^2_2(\psi^2\partial_{\overline{z}}\psi_2 +
\tilde{\psi}^2\partial_z{\tilde{\psi}}_2)\big)\Big)}\\
&\times\exp{\left[\mbox{topological terms for fermions}\right]}.
\end{split}
\end{equation}
The topological terms for the fermions are total derivatives which do 
not contribute to the fermionic equations of motion, nor do they contribute to the 
path integral. In the path integral (\ref{partosf2}) since the two fields $\psi$ and $\tilde\psi$ are decoupled, 
the partition function is the product of the Pfaffians of the 
differential operators  $\partial_z$ and $\partial_{\overline{z}}$ 
\cite{ref20},
\begin{equation}\label{partosf3}
Z = Pf(\partial_z)Pf(\partial_{\overline{z}}).
\end{equation}
As the product  $\partial_z\partial_{\overline{z}}$ is the
two-dimensional Laplacian, we get
\begin{equation}\label{partosf4}
 Z = (\det \nabla^2)^{\frac{1}{2}}.
\end{equation}
Now we impose periodicity conditions on the fermions. When translated by a period the fermions pick up a phase.
\begin{equation}\label{period}
\psi(z+\phi) = e^{2i\pi\nu}\psi(z).
\end{equation}
For $\nu \in \mathbb{Z}$, $\psi$ and $\tilde{\psi}$ are 
periodic and for $\nu \in (\mathbb{Z} + 1/2)$, both are antiperiodic. 
The combinations of periodic(P) and antiperiodic(A) boundary conditions 
for the holomorphic and antiholomorphic fields are used to define the 
spin structure of the fermions, that is, the Ramond (periodic) 
and Neveu-Schwarz (antiperiodic) sectors. 
Again, due to the factorization of the holomorphic and the antiholomorphic
parts it suffices to compute the integral for the holomorphic 
fields only and the partition function is evaluated as
\begin{equation} 
Z=|\det\partial_z|^2.
\end{equation}
Evaluating the regularized products with P and A boundary conditions,
we obtain
\begin{equation}\label{sectr1}
\begin{split}
(\det\partial_z)_{A,A} &
= \frac{\vartheta_3(\tau)}{\eta(\tau)},\qquad\mathrm{NS-NS}
\\
(\det\partial_z)_{A,P} &
= \frac{\vartheta_4(\tau)}{\eta(\tau)},\qquad\mathrm{NS-R}
\\
(\det\partial_z)_{P,A} &
= \frac{\vartheta_2(\tau)}{\eta(\tau)},\qquad\mathrm{R-NS}
\\
(\det\partial_z)_{P,P} &
= \frac{\vartheta_1(\tau)}{\eta(\tau)}.\qquad\mathrm{R-R}
\end{split}
\end{equation}
The contributions from the dual fermions are obtained in the same way with the same results.
The partition function for the fermions on $T^2$ is given by the combinations of theta functions. 
In order to obtain the partition function of type-II theories,
however, one now has to impose the GSO projections, as usual.

%The full partition function for the fermions on a torus can be written in terms of the characters of SO(2) \cite{ref21} namely:
%\begin{equation}\label{char}
%\begin{split}
%V_2 &= \frac{\vartheta_3 - \vartheta_4}{2\eta}, \\
%O_2 &= \frac{\vartheta_3 + \vartheta_4}{2\eta}, \\
%S_2 &= \frac{\vartheta_2 - i\vartheta_1}{2\eta}, \\
%C_2 &= \frac{\vartheta_2 + i\vartheta_1}{2\eta}.
%\end{split}
%\end{equation}

Upon choosing the GSO projection $\exp(i\pi F) = 1$,
we obtain the partition function of the type-IIB theory, {\it viz.}
%\begin{equation}\label{partition}
%Z^{osc}_f = \frac{(V_2 - S_2)(\overline{V}_2 - 
%\overline{S}_2)}{\eta\overline{\eta}},
%\end{equation}
\begin{equation}\label{partition2}
 Z^{osc}_f = \frac{1}{|\eta|^2}|\vartheta_3 - 
\vartheta_4 - \vartheta_2 + \vartheta_1|^2.
\end{equation}
Choosing, on the other hand, 
the GSO projection $= \exp(i\pi \tilde{F}) = (-1)^\alpha$,
with $\alpha = 1$ in the R-sector and $\alpha = 0$ in the NS-sector, 
gives the partition function of the type-IIA theory, {\it viz.}
%\begin{gather}\label{partition3}
% Z^{osc}_f = \frac{(V_2 - C_2)(\overline{V}_2 - \overline{S}_2)}{\eta\overline{\eta}},
%\\
\label{partition4}
\begin{equation} 
Z^{osc}_f = \frac{1}{|\eta|^2}(\vartheta_3 - \vartheta_4 - \vartheta_2 +
\vartheta_1)(\overline{\vartheta_3} -\overline{\vartheta_4} - 
\overline{\vartheta_2} - \overline{\vartheta_1}).
\end{equation}

Finally, combining the expressions obtained above, 
the total partition function for the type-IIB theory is obtained as
\begin{equation}\label{partition7}
Z = \prod_{i}\sum_{n_1,n_2,\omega_1,\omega_2}
\frac{1}{|\eta|^6}
\exp{\left[i\pi\tau\frac{{(p^i_L)}^2}{2} -
i\pi\overline{\tau}\frac{{(q^i_R)}^2}{2}\right]}
\times|\vartheta_3 - \vartheta_4 - \vartheta_2 + \vartheta_1|^2.
\end{equation}
The total partition function for the type-IIA theory is obtained
similarly, with the afore mentioned GSO projection as 
\begin{equation}\label{partition8}
Z = \prod_{i}\sum_{n_1,n_2,\omega_1,\omega_2}
\frac{1}{|\eta|^6}
\exp{\left[i\pi\tau\frac{{(p^i_L)}^2}{2} -
i\pi\overline{\tau}\frac{{(q^i_R)}^2}{2}\right]}
\times(\vartheta_3 - \vartheta_4 - \vartheta_2 +
\vartheta_1)(\overline{\vartheta_3} -\overline{\vartheta_4} - \overline{\vartheta_2} - \overline{\vartheta_1}).
\end{equation}
%%%
\section{Conclusion} \label{sec4}
To summarize, we have studied the supersymmetric extension of the 
doubled formalism. We evaluated the partition function of the
$\mathcal{N}=(1,1)$ NLSM on a doubled-torus,
$T^2\times \hat{T}^2$. The superstring partition functions turn out to be the squareroot of the partition function of this theory and are T-duality invariant. For the bosonic case, a topological term had to
be added to the NLSM on the doubled torus for quantum consistency. In
order to maintain supersymmetry in our case, a supersymmetric
extension of the same by fermions is required. However, these extra fermions 
conspired not to contribute to the partition functions. This supersymmetric model, thus, reproduces the
one-loop partition functions of type-II theories exactly.
Thus the calculations presented here provide a non-trivial
verification of the doubled formalism, extending to the supersymmetric
cases.
This matching, however, requires a separate choice of the 
GSO projections
as in the traditional formulations of type-II theories. 
It would be interesting to study the Hilbert space of the NLSM on the 
doubled torus and compare with the spectra of the type-II theories
directly. The fermions in the supersymmetrically extended topological terms may have an interesting role to play in such a study. Extension of this formalism
to study superstrings on more complicated spaces, such as
K3 and Calabi-Yau manifolds will also be interesting. These will be
useful in formulating a T-duality invariant string theory.
Finally, formulation presented here can be applied to the
case of $\mathcal N=(1,0)$ superspace, relevant for the Heterotic
strings.

\section{Acknowledgment}\label{sec5}
It is a pleasure to thank Sumit R. Das,
Alok Kumar, Jaydeep Majumder, 
Partha Mukhopadhyay, Koushik Ray, Sourov Roy and Soumitra Sengupta
for usefull suggestions.

\end{document}